\documentclass[draft]{agujournal2019}
\usepackage{amsmath}
\usepackage{graphicx}
\usepackage{subcaption}
\usepackage{booktabs}

\draftfalse

\journalname{Journal of Geophysical Research: Solid Earth}

\begin{document}

\title{Heterogeneous and episodic permeability evolution and correlation with
acoustic emission activity during faulting in granite}

\authors{Ado Farsi\affil{1,2,3}, Franciscus M. Aben\affil{4}, and
Nicolas Brantut\affil{1,5}}

\affiliation{1}{Department of Earth Sciences, University College London, London, UK}
\affiliation{2}{Department of Mathematics, Imperial College London, London, UK}
\affiliation{3}{Tanuki Technologies, London, UK}
\affiliation{4}{TNO Applied Geosciences, Utrecht, The Netherlands}
\affiliation{5}{Geomechanics and Scientific Drilling, GFZ Helmholtz Centre for Geosciences, Potsdam, Germany}

\correspondingauthor{Ado Farsi}{ado.farsi@tanuki.ai}

\begin{keypoints}
\item A three-dimensional inverse method estimates the local hydraulic
transmissivity in ten cells of a fault from flow and pore pressure data
\item Local transmissivity evolves episodically and heterogeneously along the
fault; the sample average grows about threefold
\item AE maps damage growth; local associations between event-count metrics
and transmissivity vary between cells and measurement intervals
\end{keypoints}

\begin{abstract}
Rock permeability changes dramatically during brittle deformation, affecting
faulting dynamics and crustal fluid flow. Here, we combine fluid-flow and local
pore-pressure measurements acquired during a quasi-static shear-failure test
in Westerly granite with a three-dimensional finite-element inverse method.
The measured flow rate and six internal pore pressures constrain the local
hydraulic transmissivities (permeability--thickness products) of ten cells
along the growing fault. Synthetic ensembles anchored to the pressure-sensor
accuracy recover these transmissivities with median mean errors of 7--9~\%,
with the fault ends generally better resolved than the interior. Applied to
the experiment, the method resolves episodic transmissivity increases,
interspersed with decreases, through fault growth and slip. Local values reach
approximately $1.6 \times 10^{-19}$~m$^3$ (16 times the bulk contribution),
and the final-stage values vary by a factor of 2.3 along the fault. Across
cells, the median absolute change in recovered transmissivity is at most
about 16~\% for $\pm 50$~\% changes in assumed fault-zone thickness. The
local one-dimensional estimates of \protect\citeA{Aben2024} agree in scale
with the three-dimensional results, with a median 3-D/apparent ratio of 0.81.
Cumulative acoustic-emission counts track the overall hydraulic evolution,
but the relations of local event density, its stagewise increments and
amplitude-weighted activity to transmissivity vary between cells. The comparison distinguishes the value of AE
for mapping damage growth from quantitative hydraulic prediction using
event-count and relative-amplitude metrics.

\end{abstract}

\section*{Plain Language Summary}
Fluids move through the Earth's crust mostly along cracks and faults, so knowing
how easily a fault transmits fluid is important for understanding earthquakes and
for storing fluids underground. Laboratory experiments can measure how much fluid
passes through a rock sample as a fault grows inside it, but such measurements
only give one average number for the whole sample: they cannot say which parts of
the fault carry the flow. In this study we combine a laboratory experiment on
granite with a computer model of the fluid flow inside the sample. Tiny pressure
sensors placed along the fault record the fluid pressure at several points, and
we use a computer method to search for the pattern of local fluid
transmissivity --- how easily fluid flows along each part of the fault --- that
best reproduces those measurements. This lets us map how fluid transport varies
from place to place along the fault, and how it changes as the fault slips. We
find that the fault's ability to transmit fluid grows episodically, with
intervening decreases, that some parts of the fault carry far more fluid than
others, and that the
places where the rock cracks most are not necessarily the places that transmit
the most fluid.

\section{Introduction}

Fluid flow in the Earth's crust is controlled by the connectivity and hydraulic
transmissivity of fractures. In bulk crystalline rock, fluids move through a
connected network of microcracks and pores, and the resulting permeability is
low, of the order of $10^{-19}$ m$^2$ or less (\protect\citeNP{Brace1968}). At
crustal scale, flow is instead focused along fault zones, which tend to be
orders of magnitude more permeable than intact rock
(\protect\citeNP{Townend2000}). Understanding how that permeability is created,
and where along a fault it resides, therefore matters for problems ranging from
the dynamics of ruptures in the presence of fluids to the management of
geothermal reservoirs and of subsurface storage.

How permeability evolves while a fault forms in intact, low porosity
crystalline rock has been characterised mainly through triaxial deformation
experiments. Deformation of initially intact granite in the brittle regime
produces a progressive increase in rock-matrix permeability, typically by a
factor of about three at stresses up to 80 \% of the failure strength
(\protect\citeNP{Zoback1975}). As failure is approached, permeability can
increase by two to three orders of magnitude (\protect\citeNP{Mitchell2008}),
closely associated with development of a shear fault that becomes the principal
flow path. This localisation of flow during shear-zone initiation is observed
in other low-porosity rocks, although its detailed evolution depends on
lithology and failure geometry (\protect\citeNP{Kluge2021}). The evolution of
permeability in granite throughout the faulting process was obtained by
\protect\citeA{Kiyama1996}, who reported a continuous increase in permeability
with increasing deformation, and in particular showed a post-peak increase of
around two orders of magnitude. The results by \protect\citeA{Kiyama1996} were
obtained at low effective pressure (5~MPa), under near-uniaxial conditions.

In all the aforementioned laboratory studies, permeability variations could
only be constrained on average, by measurements of flow entering and leaving
the sample as a whole, so that the development of flow localisation and
hydraulic heterogeneities during fault growth could not be determined in
detail. The recent development of miniature pore-pressure transducers placed
directly on a prospective fault plane makes the internal pressure field
observable (\protect\citeNP{Brantut2021FluidMethod}). Using such measurements,
\protect\citeA{Aben2024} obtained estimates of local fault transmissivity
during the faulting process. With ongoing fault slip, only modest increases in
permeability were observed (around a factor 3 on average), but local variations
were stronger and irregular. The spatio-temporal evolution of permeability was
correlated to local acoustic emission (AE) activity, which showed that fault
propagation tends to increase local permeability only when fault branches form
a connected network.

The relationship between AE activity and fluid transport properties observed by
\protect\citeA{Aben2024} has the potential to inform how microseismic activity
is linked to fluid flow patterns in the field. At field scale, hydraulic
stimulation often produces both increased injectivity and microseismicity, as
documented at the Basel enhanced-geothermal site and in hydrothermal systems
more generally (\protect\citeNP{Haring2008}; \protect\citeNP{Cox2016}). Event
locations delineate active deformation and potential flow structures.
Quantifying the associated hydraulic properties requires an additional
relationship between seismic observations and fluid flow.

Several approaches use seismicity to constrain fluid-pressure propagation.
\protect\citeA{Shapiro1997} used the migration of the seismic triggering front
at KTB to estimate hydraulic diffusivity and, with further assumptions,
permeability. \protect\citeA{Tarrahi2012} assimilated spatial event density
into a flow model in numerical experiments. At field scale,
\protect\citeA{Riffault2018} combined cumulative hypocentre density,
interpreted as a pressure-rise proxy, with injectivity and a reservoir model to
infer permeability enhancement at Paralana. These methods use the spatial and
temporal distribution of events through a flow model; they do not require each
local event to create a prescribed permeability increment.

Other approaches use the amount or rate of seismic deformation.
\protect\citeA{RothertBaisch2010} mapped cumulative seismic shear slip to
identify hydraulically connected regions at Cooper Basin.
\protect\citeA{Fang2018} mapped permeability at Newberry from event magnitudes
through a fracture-aperture model involving stress drop, elastic properties and
dilation angle. For EGS Collab and Utah FORGE, \protect\citeA{Yu2024} used
two-minute event rates and a cumulative sum of logarithmic seismic moments as
inputs to a statistical model of permeability evolution; they also proposed a
scaling relationship between permeability increase and event magnitude, based
on a simple mechanical model whereby fault slip is assumed proportional to
fault zone opening. While intuitively attractive, such scaling relationships
rely on a number of simplifications and assumptions that need further testing,
ideally using independent local estimates of transport properties in
conjunction with seismic data. Here, we propose to determine the co-evolution
of microseismic activity and fluid transport properties during faulting in a
laboratory rock sample tested under upper crustal conditions. We leverage the
dataset obtained by \protect\citeA{Aben2024} and go beyond their initial
analysis by extracting the detailed, three-dimensional evolution of local
permeability during fault formation and growth in Westerly granite, and
comparing it to AE statistics. The key methodological step is the development
of a three-dimensional finite-element inverse method to estimate the local
hydraulic transmissivity in ten cells along the fault at multiple stages of
deformation. Inverse methods of this kind, sometimes called ``permeability
tomography'', are well established for aquifer permeability at the scale of
boreholes and reservoirs (e.g., \protect\citeNP{GottliebDietrich1995};
\protect\citeNP{Butler1999}; \protect\citeNP{YehLiu2000}), but they have not
yet been applied to laboratory-scale specimens under elevated confining
pressure, because the necessary internal measurements did not exist. We
systematically test the results and sensitivity of our new method, and compare
the recovered transmissivity with the spatial distribution of AEs to relate
local hydraulic evolution to accumulated damage. Specifically, we ask whether
cumulative local AE event density or corrected-amplitude density, or their
increments between hydraulic measurements, provide a common quantitative
relation to local transmissivity or its change.

\section{Methods}

\subsection{Experimental data}
The data used here are those of \protect\citeA{Aben2024}; we summarise only what is
needed to define the inverse problem, and refer to that paper for the full
experimental procedure.

\begin{figure}
    \centering
    \includegraphics[width=\linewidth]{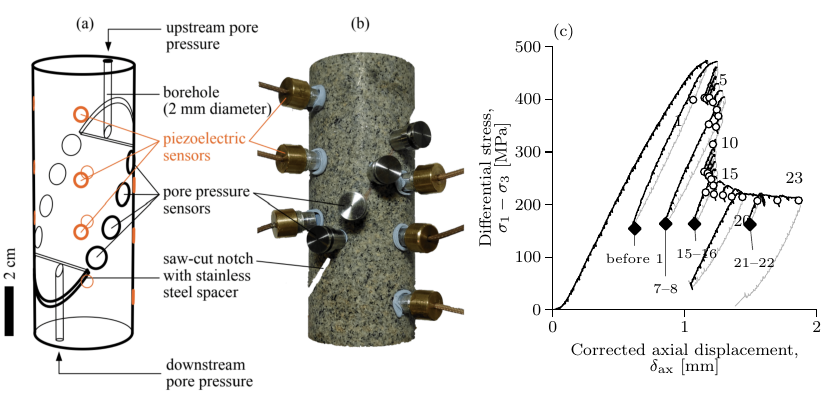}
    \caption{Experimental configuration and loading history. (a) Schematic of
    the sample, notches, boreholes, pore-pressure sensors and piezoelectric
    sensors. (b) Photograph of the instrumented sample before jacketing; panels
    (a) and (b) are from \protect\citeA{Aben2024}. (c) Differential stress as a
    function of corrected axial displacement, regenerated from the experimental
    data supplied with this study. The complete loading--unloading history is
    light grey, increasing-load branches are black, and open circles mark the
    23 stages at which deformation was paused and steady flow was imposed. A
    subset of stage numbers is shown for orientation. Diamonds mark the
    overnight holds before stage 1 and between stages 7--8, 15--16 and 21--22.}
    \label{fig:experimental_overview}
\end{figure}

A 40~mm diameter, 100~mm long Westerly granite cylindrical sample was prepared
with two notches, cut at 30 degrees from the loading axis, to enforce a
predictable shear-localisation geometry. The sample was instrumented with eight
miniature pore-pressure transducers (gauges $P_1$--$P_8$) arranged along the
prospective fault and with an array of 12 piezoelectric transducers used to
locate acoustic emissions. The geometry and sensor arrangement are shown in
Figures~\ref{fig:experimental_overview}a and
\ref{fig:experimental_overview}b.

The sample was confined at 85~MPa and deformed at a constant shortening rate,
with the pore pressure maintained at 5~MPa. Deformation was paused at 23 stages
(Figure~\ref{fig:experimental_overview}c); at each pause, the differential stress was decreased (to prevent further
deformation), the downstream pore-pressure line was vented, and the flow rate and internal pore pressures
were recorded once steady flow was established. Before stage 1 and between
stages 7 and 8, 15 and 16, and 21 and 22, the experiment was stopped:
differential stress was brought down to around 150~MPa or lower, and all
stresses and pressures were held constant for a period of around 14.5~hours
(overnight). Deformation was then resumed as before. Deformation is reported
below as the equivalent fault slip $s$, which is defined as the average
inelastic shortening projected
onto the slip direction, referenced to the displacement at peak stress
\protect\cite{Aben2024}. Gauge $P_1$ suffered from substantial,
irregular drift throughout the experiment, and $P_4$ developed comparable
drift after stage 15 \protect\cite{Aben2024}. Both records are excluded from
every inversion, so six of the eight internal pressures ($P_2$, $P_3$,
$P_5$--$P_8$) are used as observations.

\subsection{Forward problem}
The forward problem is steady, incompressible single-phase flow, for which mass
conservation and Darcy's law reduce to
\begin{equation}
    \nabla \cdot \mathbf{u} = 0,
    \label{eq:conservation_equation}
\end{equation}
\begin{equation}
    \mathbf{u} = -\frac{\mathbf{k}}{\mu} \nabla p,
    \label{eq:darcy_law}
\end{equation}
where $\mathbf{k}$ is the permeability tensor, $\mu$ the fluid viscosity, $p$ the
fluid pressure and $\mathbf{u}$ the Darcy velocity. Gravity is neglected.

Equations~\eqref{eq:conservation_equation} and~\eqref{eq:darcy_law} are solved in
mixed form \protect\cite{Auricchio2017MixedMethods} with the finite element library
Firedrake \protect\cite{Ham2023FiredrakeManual}. The weak form reads
\begin{equation}
    \int_{\Omega} w \, \nabla \cdot \mathbf{u} \, \mathrm{d}\Omega = 0,
    \label{eq:weak_conservation_equation}
\end{equation}
\begin{equation}
    \int_{\Omega} \mathbf{K}^{-1}\mathbf{u} \cdot \mathbf{v} \, \mathrm{d}\Omega
    - \int_{\Omega} p \, \nabla \cdot \mathbf{v} \, \mathrm{d}\Omega
    = - \int_{\Gamma} p \, \mathbf{v} \cdot \mathbf{n} \, \mathrm{d}\Gamma,
    \label{eq:weak_darcy_law}
\end{equation}
with $\mathbf{K}=\mathbf{k}/\mu$, where $\mathbf{v}$ and $w$ are test functions,
$\Omega$ is the sample volume and $\Gamma$ its boundary. The velocity is
discretised with Raviart--Thomas elements of degree two and the pressure with
discontinuous Lagrange elements of degree one.

The computational mesh reproduces the sample geometry: a
40~mm diameter, 100~mm long cylinder containing two 15~mm deep notches inclined
at 30 degrees to the axis and connected to the sample ends by 2~mm diameter
boreholes. It was generated with Gmsh \protect\cite{Geuzaine2009} with an average element
size of 2.5~mm. The notch volumes are meshed separately so that they can be
assigned their own properties.

The permeability of the bulk rock, $K_0$, was taken as $10^{-18}$~m$^2$. The
sample-average apparent permeability measured just before fault growth, at
peak stress, is $7.5 \times 10^{-19}$~m$^2$ \protect\cite{Aben2024}, consistent
with values reported for Westerly granite at stresses of 80--90~\% of peak
strength \protect\cite{Zoback1975, Mitchell2008}. This is a prescribed
off-fault value that we assume constant. The dependence of the results on
the specific choice of $K_0$ will be tested by repeating inversions with
different values of this parameter (Supporting Information Text~S3). The notches are filled with a
stainless steel spacer that leaves a gap around the insert, so the fluid moves
freely within them; they were assigned a permeability three orders of magnitude
larger than $K_0$, which makes the pore pressure within a notch essentially
uniform and equal to the pressure imposed at the corresponding borehole.

Flow is driven by pressure boundary conditions applied on the flat ends of the
cylinder and on the inner surfaces of the two boreholes, with a no-flow condition
on the lateral surface. All quantities are non-dimensionalised, with the pressure
normalised as $\tilde{p} = (p - p_\mathrm{down})/(p_\mathrm{up} -
p_\mathrm{down})$, so that the imposed boundary pressures are $1$ upstream and
$0$ downstream, lengths by $L_\mathrm{ref}=0.1$~m, and permeabilities by
$K_\mathrm{ref}=10^{-17}$~m$^2$. The surface-integrated flow rate is normalised
by $q_\mathrm{ref}=K_\mathrm{ref}\Delta p L_\mathrm{ref}/\mu$, equal to
$4.988\times10^{-9}$~m$^3$~s$^{-1}$, with $\Delta p=4.988$~MPa and
$\mu=10^{-3}$~Pa~s. The same normalisation is applied to the measured
pressures and flow rate before they enter the inverse problem, so that model
and data are compared as dimensionless quantities.

\subsection{Inverse problem}
\label{sec:inverse_problem}
\begin{figure}
    \centering
    \begin{subfigure}{0.46\textwidth}
        \includegraphics[width=\linewidth]{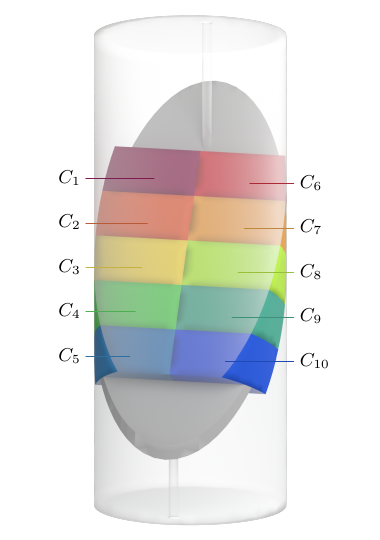}
        \caption{}
        \label{fig:inverse_problem_unknowns}
    \end{subfigure}
    \begin{subfigure}{0.46\textwidth}
        \includegraphics[width=\linewidth]{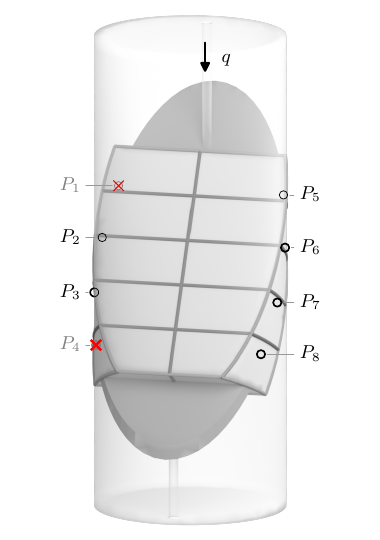}
        \caption{}
        \label{fig:inverse_problem_observables}
    \end{subfigure}
    \caption{Unknowns (a) and observables (b) of the inverse problem. (a) The
    fault zone is discretised into ten cells $C_1$--$C_{10}$; each cell
    carries a uniform fault-zone permeability $k_f^i$, and the ten $k_f^i$
    are the unknowns of the inverse problem. (b)
    The observables: the six retained internal pore pressures ($P_2$, $P_3$,
    $P_5$--$P_8$; the excluded $P_1$ and $P_4$ are crossed out) and the flow
    rate $q$ through the boreholes (arrow).}
    \label{fig:inverse_problem}
\end{figure}

A 10~mm thick layer of the sample between the two notches is taken as the fault
zone. Its thickness was chosen from the region of highest AE density and from
post-test visual inspection of the fault traces documented in Figures~6 and 9 of
\protect\citeA{Aben2024}. The fault zone is divided into ten
cells, five along the fault dip and two across it, named $C_1$--$C_{10}$
(Figure~\ref{fig:inverse_problem_unknowns}), with the five divisions
positioned so that the pressure gauges fall at the interfaces between
cells: the gauge and model interfaces agree to within 1~mm. The simulated
pressures are in every case evaluated at the true gauge coordinates. Cells
$C_1$--$C_5$ lie on the Western side carrying gauges $P_1$--$P_4$, and cells
$C_6$--$C_{10}$ on the Eastern side carrying gauges $P_5$--$P_8$; numbering
increases from the upstream (South) to the downstream (North) notch. The
mesh-centroid and gauge-coordinate check confirms this orientation.

Away from the fault zone the permeability is held at $K_0$ as differential
stress decreases from peak to residual strength
(Figure~\ref{fig:experimental_overview}c). We assume that the bulk permeability
increase produced by the crack network near peak stress is largely retained
during this unloading. Unfortunately, the sensor and sample geometries do
not allow us to test directly the validity of this assumption: such a test
would require additional off-fault sensors that could not be fitted with our
setup. Nevertheless, the use of an approximately constant bulk permeability
during unloading is supported by cyclic-loading experiments on Westerly
granite, which show substantial permeability
hysteresis: much of the damage-induced increase is retained after unloading,
and the attained permeability is controlled primarily by the maximum stress
previously experienced \protect\cite{Zoback1975,Mitchell2008}. Within cell
$C_i$ the fault-zone permeability takes the uniform value
$k_f^i$, and the ten permeabilities $k_f^i$ are the unknowns of the inverse
problem, bounded by $K_0 \leq k_f^i \leq 1000K_0$ for the post-peak stages.
The observations are
the six retained internal pressures and the flow rate $q$
(Figure~\ref{fig:inverse_problem_observables}).

Because the fault zone is a thin, relatively permeable layer carrying flow
along itself, the observables are, to leading order, sensitive to the product
of its permeability and its thickness rather than to the two separately: the
quantity the data constrain in each cell is the fault-parallel fluid
transmissivity $\langle k_f t \rangle_i = k_f^i\, t$, in the notation of
\protect\citeA{Aben2024}, with $t$ the assumed
fault-zone thickness. The results are therefore reported as transmissivities;
Section~\ref{sec:thickness_sensitivity} quantifies how far this equivalence
holds. The off-fault permeability $K_0$ controls the contribution from the
surrounding rock and is prescribed separately from the fault-zone thickness.

The permeabilities are estimated by least squares minimisation. Writing
$\mathbf{k}_f = (k_f^1, \ldots, k_f^{10})$ for the vector of unknowns,
$\tilde{q}$ for the
normalised flow rate, $\tilde{p}_g$ for the normalised pressure at gauge $g$, and
using a circumflex for the measured values, the objective function is
\begin{equation}
    J(\mathbf{k}_f) =
    \left( \frac{\tilde{q}(\mathbf{k}_f) - \hat{\tilde{q}}}{\hat{\tilde{q}}} \right)^{2}
    + \sum_{g \in \mathcal{G}}
    \left( \frac{\tilde{p}_g(\mathbf{k}_f) - \hat{\tilde{p}}_g}{\hat{\tilde{p}}_g} \right)^{2},
    \label{eq:objective_function}
\end{equation}
where $\mathcal{G}$ is the set of six retained gauges. Each residual is normalised
by its own measured value, so that every term of
Equation~\eqref{eq:objective_function} is a relative error and the flow rate and
the pressures contribute with equal weight despite being different physical
quantities; no further weighting factors are introduced. The simulated pressure
$\tilde{p}_g$ is obtained by point evaluation of the computed pressure field at
the gauge location.

The gradients of $J$ with respect to the ten permeabilities are computed with an
adjoint method, using Pyadjoint \protect\cite{Mitusch2019Dolfin-adjointFiredrake}, which
makes the cost of the gradient independent of the number of parameters. The
minimisation uses the limited-memory bounded Broyden--Fletcher--Goldfarb--Shanno algorithm
\protect\cite{Fletcher2020PracticalEdition}. For the first stage the permeabilities are
initialised at $k_f^i = K_0$, i.e.\ at the bulk value; each subsequent stage is initialised
from the optimised permeabilities of the previous stage.

\subsection{Acoustic emission event density field}
The AE recorded during the experiment were used to build an AE event density
field for each stage. Each located event was represented by a Gaussian
distribution centred on its location, with a standard deviation of one third
of the 2.5~mm location accuracy, i.e.\ about 0.8~mm \protect\cite{Aben2024}. The AE event density field at a given stage is the superposition
of the Gaussians of all events recorded up to the end of that stage. The field
was interpolated onto the mesh with a scalar polynomial function space of degree
two.

The AE density field is not used as an input to the permeability inversion. It is used only,
after the fact, as an independent description of where damage accumulates, and to
define a local AE density for each cell of the fault zone.

To distinguish accumulated damage from recent activity, we compare, in each cell, both the
cumulative AE density and its increment between successive hydraulic
measurements with transmissivity. Changes in transmissivity are expressed as
$\log_{10}(T_j/T_{j-1})$, where $T=\langle k_f t\rangle$ and $j$ indexes the
23 post-peak measurements, giving 22 paired intervals. We also calculate
sample-wide catalogue event rates over the 120~s preceding the last located
event before each measurement pause, with 60 and 300~s sensitivity windows.
Supporting Information Text~S4 defines these count-based metrics and reports
descriptive Spearman correlations. To include event size, we use the
AE amplitudes $A_e$ (in volts), corrected for source--receiver distance following
\protect\citeA{Zang2000}, and define $M_{\mathrm{AE},e}=\log_{10}A_e$.
Using events with amplitudes, we compare
unweighted and amplitude-weighted local activity, and the local mean
$M_{\mathrm{AE}}$, with transmissivity changes (Supporting Information
Text~S5 and Table~S5).
For the cumulative comparison, we accumulate corrected amplitudes
$A_e=10^{M_{\mathrm{AE},e}}$ up to each measurement, using the same
spatial averaging and the same events for counts and amplitudes.

\section{Results}

\subsection{Transmissivity structure and its evolution}
The flow test acquired before deformation provides an intact-rock reference.
Using its homogeneous flow calibration of $2.09\times10^{-20}$~m$^2$ for
the surrounding rock, a ten-cell inversion recovers local permeabilities of
$4.98\times10^{-21}$--$3.72\times10^{-20}$~m$^2$ and closely fits the
measured flow and six pressures (Supporting Information Text~S2 and
Table~S2). This provides a spatial reference at the much lower intact-rock
permeability scale.

\begin{figure}
    \centering
    \begin{subfigure}[b]{0.24\textwidth}
        \includegraphics[width=\linewidth]{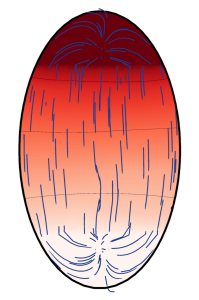}
        \caption{$s = 0.15$~mm}\label{fig:results_stage_2_projection}
    \end{subfigure}
    \hfill
    \begin{subfigure}[b]{0.24\textwidth}
        \includegraphics[width=\linewidth]{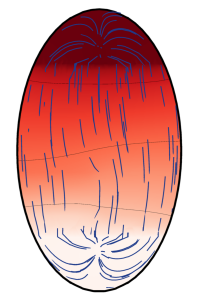}
        \caption{$s = 0.60$~mm}\label{fig:results_stage_15_projection}
    \end{subfigure}
    \hfill
    \begin{subfigure}[b]{0.24\textwidth}
        \includegraphics[width=\linewidth]{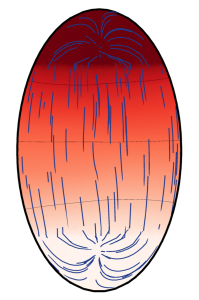}
        \caption{$s = 1.41$~mm}\label{fig:results_stage_23_projection}
    \end{subfigure}
    \hfill
    \begin{subfigure}[b]{0.096\textwidth}
        \includegraphics[width=\linewidth]{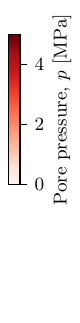}
    \end{subfigure}

    \begin{subfigure}{\textwidth}
        \includegraphics[width=\linewidth]{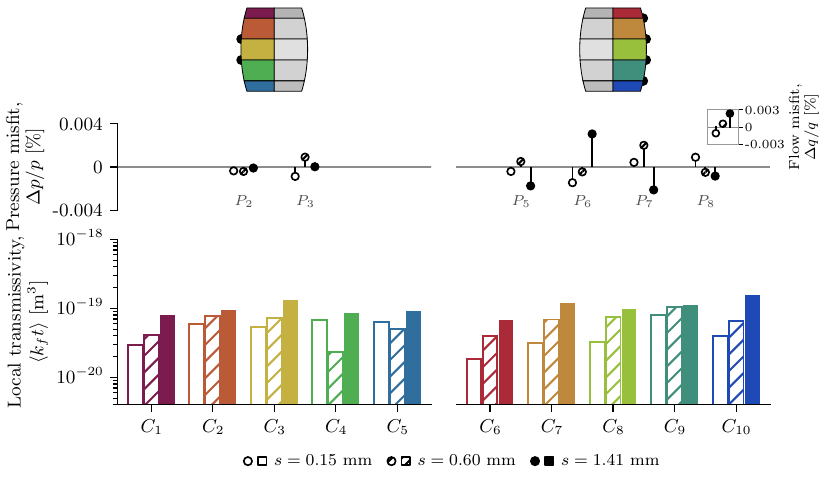}
        \caption{}\label{fig:results_stages_fit}
    \end{subfigure}
    \caption{Results at equivalent slips $s=0.15$~mm (early post-peak),
    $0.60$~mm (roughly halfway through the experiment) and $1.41$~mm
    (the last measurement), obtained by
    solving the forward problem once with the permeabilities recovered for
    each measurement. (a--c) Pore pressure on the fault plane at
    $s=0.15$~mm (a), $0.60$~mm (b) and $1.41$~mm (c), with the flow lines
    through the sample (blue); the three panels
    share the colour scale. (d) Top: relative misfit of the recovered pore
    pressure at the six working gauges (open circles: $s=0.15$~mm; hatched:
    $0.60$~mm; filled: $1.41$~mm), each drawn at the interface between the
    cells it separates; the black half-dots on the fault miniatures mark
    the gauge positions; the inset gives the relative misfit of the flow rate.
    The expanded misfit scales resolve the small numerical residuals; they
    do not represent measurement uncertainty.
    Bottom: recovered local transmissivity of each cell (open bars:
    $s=0.15$~mm; hatched: $0.60$~mm; filled: $1.41$~mm), in the colours of
    Figure~\ref{fig:inverse_problem_unknowns}.}
    \label{fig:results_stages}
\end{figure}

\begin{figure}
    \centering
    \includegraphics[width=\linewidth]{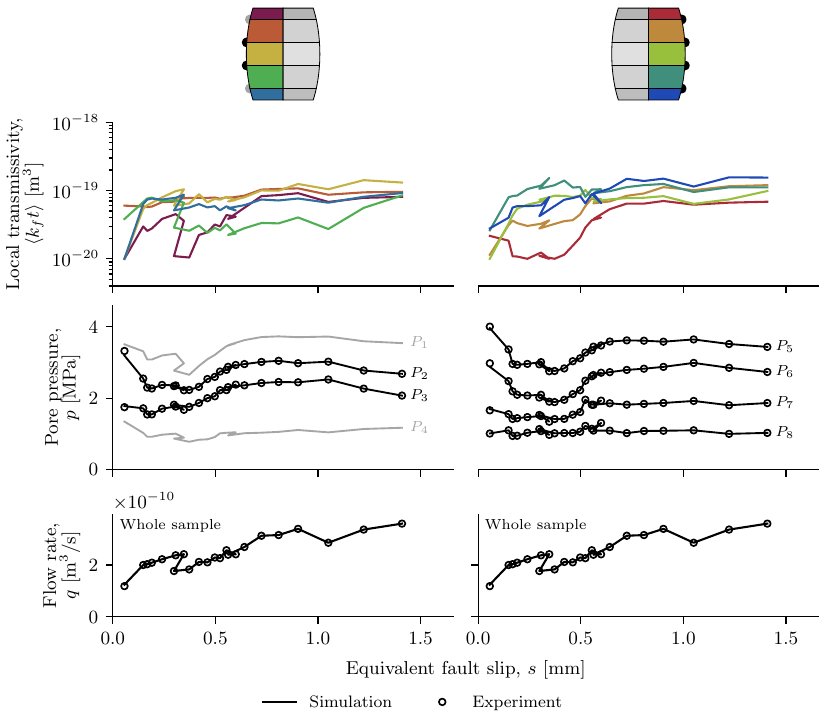}
    \caption{Recovered structure and fit to the data for all 23 measurements,
    against equivalent fault slip. The miniatures mark the side of the fault
    each column shows (left: cells $C_1$--$C_5$; right: cells $C_6$--$C_{10}$), with
    black half-dots marking the working gauges (as in
    Figure~\ref{fig:results_stages}) and light-grey ones the excluded gauges
    $P_1$ and $P_4$. Top row: local transmissivity $\langle k_f t \rangle$ of
    each cell, in the colours of
    Figure~\ref{fig:inverse_problem_unknowns}. Middle row: simulated (black
    lines, named at their right ends) and measured (open circles) pore
    pressure at each side's gauges; measurements from $P_1$ and $P_4$ are not
    used or plotted, and their simulated curves are drawn in the same light
    grey as their dots. Bottom row: simulated and measured flow rate, a single observable for the
    whole sample. This comparison is repeated beneath both columns to maintain
    the same slip-axis scale throughout. Lines connect measurements in chronological order; their two
    short backward segments on the horizontal axis reflect the unload--reload
    slip reversals from $0.35$ to $0.30$~mm and from $0.60$ to $0.56$~mm.}
    \label{fig:optimisation_results}
\end{figure}

Applying the inversion to all 23 stages gives the evolution of the ten local
transmissivities through fault growth and slip.
Figure~\ref{fig:results_stages} illustrates three stages: ``2'', shortly after
peak stress, ``15'', roughly halfway through the experiment, and ``23'', the
final stage. The recovered combinations of transmissivities match both the measured
flow rate and the six measured pressures, and the pressure fields show how the
heterogeneity reorganises the flow. The fit is equally close at stages 2--23
(Figure~\ref{fig:optimisation_results}), with maximum pressure and flow
residuals of 0.0054 and 0.0027~\%, respectively. At stage 1, four cells reach
the prescribed background permeability and the corresponding residuals are
3.90 and 4.83~\%; using $K_0=7.5\times10^{-19}$~m$^2$ also fits this first
stage closely (Supporting Information Text~S3).

Figure~\ref{fig:optimisation_results} shows the recovered local
transmissivities against equivalent fault slip. With increasing deformation and
fault growth, the fault-volume-weighted mean transmissivity rises from
$2.38\times10^{-20}$~m$^3$ at stage 1 ($s=0.06$~mm) to
$7.93\times10^{-20}$~m$^3$ at stage 7 ($s=0.35$~mm). It falls by 31~\%
across the following unload--reload transition, then fluctuates without a
consistent trend up to $s=0.64$~mm. A later increase reaches
$9.83\times10^{-20}$~m$^3$ at $s=0.90$~mm, followed by a 19~\% decrease
and recovery to $1.05\times10^{-19}$~m$^3$ at $s=1.41$~mm. The development
of transmissivity is episodic rather than continuous.

The recovered structure is heterogeneous. The cell near the downstream notch
on the gauge 5--8 side ($C_{10}$) has the highest transmissivity from
$s=0.64$~mm onwards in the primary inversion, reaching approximately
$1.6\times10^{-19}$~m$^3$ (16 times the bulk contribution $K_0t$).
The largest within-stage contrast is 15.3 at stage 7, where $C_6$ reaches
the lower bound; at the final stage the local values span
$6.85\times10^{-20}$--$1.55\times10^{-19}$~m$^3$, a factor of 2.3.
The permeability equivalents of these values assume the nominal 10~mm
fault-zone thickness; Section~\ref{sec:thickness_sensitivity} quantifies how
the recovered quantities depend on it.

\subsection{Comparison with the local apparent estimates}
\label{sec:one_dimensional_comparison}
We compared the three-dimensional results with the local apparent
transmissivities of \protect\citeA{Aben2024}, which were estimated for
the same ten regions at every stage using (their Equations~5 and 6)
\begin{equation}
    \langle k_f t\rangle_i^\mathrm{app}
    = \frac{q\mu\Delta X_i}{\langle\ell\rangle_i\,\Delta p_i},
    \label{eq:apparent_one_dimensional_transmissivity}
\end{equation}
where $q$ is the measured total flow rate, $\mu$ the water viscosity,
$\Delta X_i$ the down-dip distance between bounding pressure measurements or
notch tips, $\langle\ell\rangle_i$ the mean elliptical fault width, and
$\Delta p_i$ the local pressure drop. This calculation assumes uniform flow
allocation, whereas the three-dimensional inversion includes off-fault and
cross-fault flow.

Across the 214 local comparisons unaffected by late $P_4$ drift, the median
3-D/apparent transmissivity ratio is 0.81, with a 5th--95th percentile range
of 0.44--1.92. The two approaches therefore agree in scale, while the
three-dimensional calculation resolves differences in local flow allocation.
The comparison method and stage-by-stage summary are reported in
Supporting Information Text~S1 and Table~S1.

\subsection{Verification and sensitivity of the inverse method}
\label{sec:verification}
The ability of the method to recover a known transmissivity structure was tested
against ensembles of synthetic data. Synthetic datasets were generated by
solving the forward problem with a prescribed transmissivity structure and
perturbing the resulting pressures with seeded Gaussian noise; as for the
experimental data, gauges $P_1$ and $P_4$ are excluded, so each synthetic
inversion uses the same seven observations as the experiment. Two target
structures were used: one at the configuration recovered at stage 15,
roughly halfway through the experiment, and one at stage 23, the last stage.
In each case the target permeabilities were the recovered values, and the
inversion was started from the values recovered for the preceding stage.

Two noise models were used, with 20 realisations of each nonzero level. The first perturbs every gauge in proportion to its own
reading, at 1, 3 and 5~\%. For each realisation, we generated one set of
random pressure errors and scaled that same set to the three noise levels.
Comparisons between levels therefore change the error amplitude while
retaining the same pattern across gauges. The second uses a constant Gaussian standard deviation of
0.06~MPa, anchored to the worst-case accuracy of the gauges
\protect\cite{Aben2024}; relative to the individual gauge readings on the 5~MPa
pressure drop across the sample this amounts to approximately 2--6~\%.
This provides a conservative test at the scale of the instrumental accuracy. The measured flow
rate is left unperturbed. In the absence of noise the transmissivities are
recovered with mean relative errors below 0.01~\% for both structures.
Table~\ref{tab:synthetic_errors} reports the median and the 5--95~\% band,
over the 20 realisations, of the mean and maximum relative error of the ten
transmissivities.

\begin{figure}
    \centering
    \includegraphics[width=\linewidth]{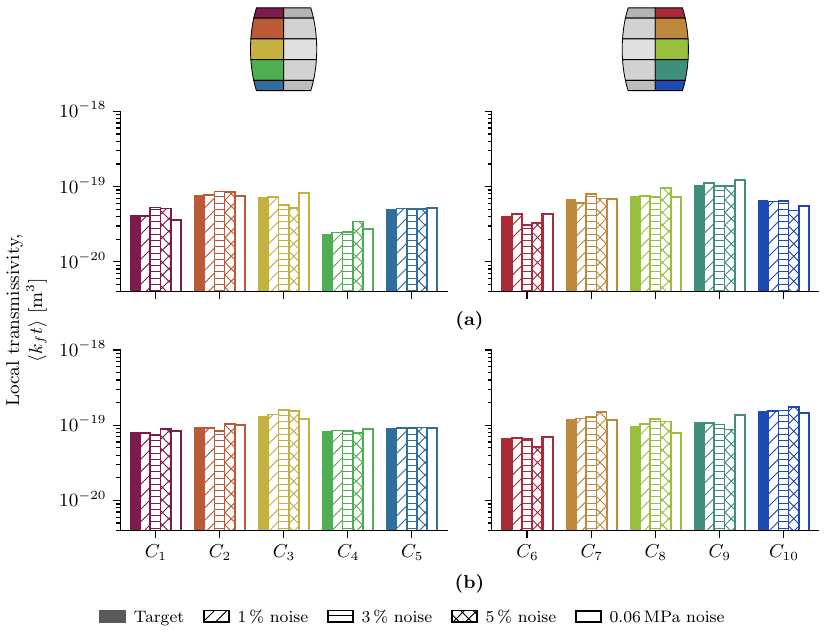}
    \caption{Recovery of two transmissivity structures from synthetic data:
    (a) the structure recovered at stage 15 and (b) that at stage 23. The miniatures at the top mark the
    side of the fault each column shows (left: cells $C_1$--$C_5$; right:
    cells $C_6$--$C_{10}$). Each cell carries five bars: the target local
    transmissivity $\langle k_f t \rangle$ and the values recovered at 1, 3 and 5~\%
    noise proportional to each gauge reading and with the constant absolute
    noise scale of 0.06~MPa anchored to the accuracy of the gauges. Each
    recovered bar is the ensemble member closest to the median error of the 20
    noise realisations of that level (Table~\ref{tab:synthetic_errors}). The
    logarithmic axis uses the same limits as the other transmissivity panels.}
    \label{fig:validation_synthetic}
\end{figure}

Three features stand out. First, the error grows monotonically with the noise
level in every paired realisation, and the two structures are recovered
comparably well at matched noise: at 1~\% the median mean error (the median
across realisations of the mean error across the ten cells) is 3.8~\% for
the stage-15 structure and 2.7~\% for stage 23. Second, at the
instrument-anchored level the median mean error is 9.1~\% for stage 15 and
7.5~\% for stage 23. Third, errors are generally larger in the interior cells
on the gauge 5--8 side, while the fault ends are better recovered.
Figure~\ref{fig:validation_synthetic} shows, for both structures and every
noise model, the realisation whose mean error is closest to the ensemble
median.

\begin{table}
    \caption{Recovery of the ten transmissivities $\langle k_f t \rangle$ from
    synthetic data: median and [5--95~\%] band, over 20 seeded noise
    realisations, of the mean and maximum relative error across the ten
    cells, for the stage-15 and stage-23 target structures. Percentage
    noise is applied in proportion to each gauge reading; the 0.06~MPa row
    uses a constant Gaussian standard deviation anchored to the gauge accuracy
    \protect\cite{Aben2024}, equivalent to 2--6~\% of the gauge readings. The
    0~\% row is the noise-free recovery error. One stage-23 realisation at
    5~\% noise has a 257~\% mean cell error and remains in the ensemble;
    the reported bands are ensemble quantiles.}
    \label{tab:synthetic_errors}
    \centering
    \small
    \begin{tabular}{lcccc}
    \toprule
    & \multicolumn{2}{c}{Stage 15 structure} &
      \multicolumn{2}{c}{Stage 23 structure} \\
    \cmidrule(lr){2-3} \cmidrule(lr){4-5}
    Noise & Mean [\%] & Max [\%] & Mean [\%] & Max [\%] \\
    \midrule
    0~\% & 0.004 & 0.009 & 0.008 & 0.022 \\
1~\% & 3.8 [1.9--6.1] & 9.1 [5.3--16.7] & 2.7 [1.9--5.9] & 7.3 [4.1--12.9] \\
3~\% & 11.0 [6.0--21.4] & 27.8 [17.0--57.3] & 8.3 [5.8--18.5] & 21.5 [13.3--48.0] \\
5~\% & 18.5 [10.4--47.6] & 46.4 [28.4--122.2] & 14.7 [9.8--45.8] & 37.2 [23.8--157.4] \\
0.06~MPa & 9.1 [4.6--14.4] & 20.0 [11.2--34.8] & 7.5 [5.5--14.4] & 18.6 [12.2--28.8] \\%
\bottomrule
    \end{tabular}
\end{table}

\begin{figure}
    \centering
    \includegraphics[width=\linewidth]{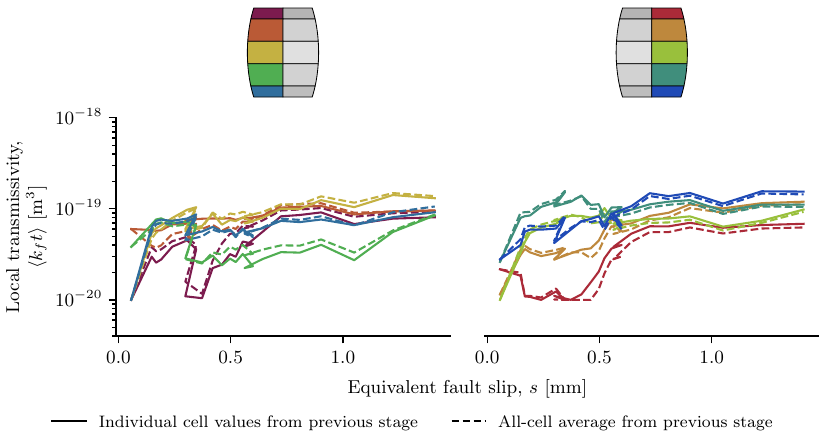}
    \caption{Sensitivity of the optimised local transmissivities
    $\langle k_f t \rangle$ to the
    initial guess. Continuous lines: each stage initialised from the previous
    stage's individual optimised cell values; dashed lines: from their
    all-cell average. The miniatures mark the side of the fault each column
    shows (left: cells $C_1$--$C_5$; right: cells $C_6$--$C_{10}$).}
    \label{fig:sensitivity_optimised_ci}
\end{figure}

As an additional check, the sensitivity of the result to the initial guess was
tested by repeating the inversion of every stage from a different starting point:
instead of the previous stage's individual optimised cell values, their
all-cell average was used. Figure~\ref{fig:sensitivity_optimised_ci} compares the two
sets of results. The median absolute difference is 9.3~\% across cells and
2.6~\% for the fault-volume-weighted mean, supporting a consistent
fault-scale history across these initial guesses.

\begin{figure}
    \centering
    \includegraphics[width=\linewidth]{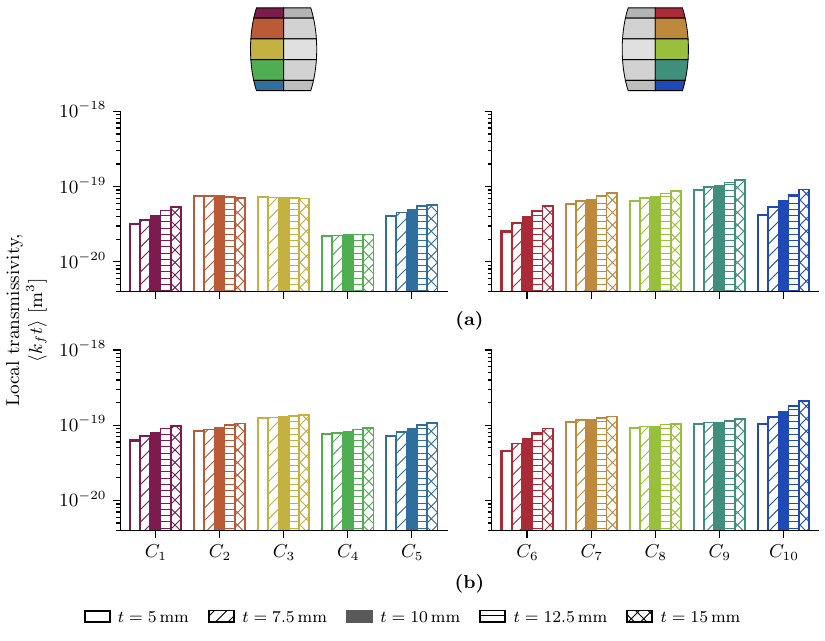}
    \caption{Local transmissivity $\langle k_f t \rangle$ recovered from
    the experimental data when the inversion assumes a fault-zone thickness
    $t$ of 5, 7.5, 10, 12.5 or 15~mm (one bar per assumption, the nominal
    10~mm solid), at (a) stage 15 ($s = 0.60$~mm) and (b) stage 23
    ($s = 1.41$~mm). The miniatures mark the side of the fault each column
    shows (left: cells $C_1$--$C_5$; right: cells $C_6$--$C_{10}$). The
    bars within each cell show the sensitivity of the recovered
    transmissivities to the assumed thickness, while the corresponding
    permeabilities scale roughly inversely with it. Cell-volume ratios account
    for clipping of the layer at the cylindrical boundary.}
    \label{fig:thickness_sensitivity_bars}
\end{figure}

Finally, we test the sensitivity to the assumed fault-zone
thickness.\label{sec:thickness_sensitivity} The nominal 10~mm is an
interpretive choice, based on the region of highest AE density and on
post-test inspection. Repeating the full 23-stage inversion with the
fault-zone layer meshed at 5, 7.5, 10, 12.5 and 15~mm fits stages 2--23
equally closely (objective values below $7.1\times10^{-9}$); stage 1 reaches
the prescribed background limit at each thickness. Across the five inversions with varying fault thickness, the recovered permeabilities rescale roughly inversely with the assumed
thickness (Figure~\ref{fig:thickness_sensitivity_bars}), whereas the median
absolute changes in recovered transmissivity across all 230 cell--stage
estimates are 15.4, 7.7, 8.3 and 15.8~\% for assumed thicknesses of 5, 7.5,
12.5 and 15~mm, respectively. Thus, for $\pm 50$~\% changes in assumed
fault-zone thickness, the median absolute change in recovered transmissivity
is at most about 16~\%. The degeneracy is not exact, because the finite
layer exchanges fluid with the surrounding rock, but transmissivity is the
better-conditioned quantity.

\subsection{Acoustic emission density and its relation to transmissivity}
\begin{figure}
    \centering
    \begin{minipage}[c]{0.86\textwidth}
        \centering
        \begin{subfigure}[b]{0.239\linewidth}
            \includegraphics[width=\linewidth]{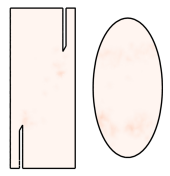}
            \caption{$s = 0.06$~mm}
        \end{subfigure}\hfill
        \begin{subfigure}[b]{0.239\linewidth}
            \includegraphics[width=\linewidth]{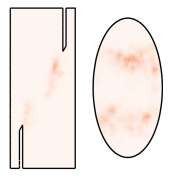}
            \caption{$s = 0.19$~mm}
        \end{subfigure}\hfill
        \begin{subfigure}[b]{0.239\linewidth}
            \includegraphics[width=\linewidth]{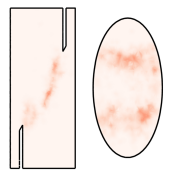}
            \caption{$s = 0.35$~mm}
        \end{subfigure}\hfill
        \begin{subfigure}[b]{0.239\linewidth}
            \includegraphics[width=\linewidth]{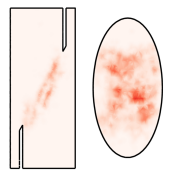}
            \caption{$s = 0.46$~mm}
        \end{subfigure}

        \vspace{4pt}
        \begin{subfigure}[b]{0.239\linewidth}
            \includegraphics[width=\linewidth]{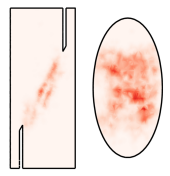}
            \caption{$s = 0.60$~mm}
        \end{subfigure}\hfill
        \begin{subfigure}[b]{0.239\linewidth}
            \includegraphics[width=\linewidth]{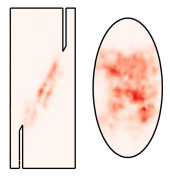}
            \caption{$s = 0.72$~mm}
        \end{subfigure}\hfill
        \begin{subfigure}[b]{0.239\linewidth}
            \includegraphics[width=\linewidth]{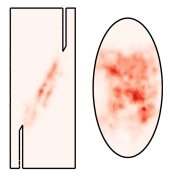}
            \caption{$s = 1.05$~mm}
        \end{subfigure}\hfill
        \begin{subfigure}[b]{0.239\linewidth}
            \includegraphics[width=\linewidth]{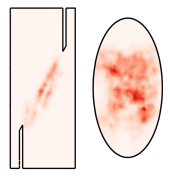}
            \caption{$s = 1.41$~mm}
        \end{subfigure}
    \end{minipage}%
    \begin{minipage}[c]{0.096\textwidth}
        \centering
        \includegraphics[width=\linewidth]{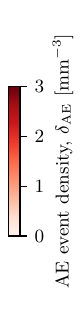}
    \end{minipage}
    \caption{Cumulative AE event density field at eight representative stages ---
    (a)--(h) correspond to stages 1, 4, 7, 11, 15, 18, 21 and 23 of the
    experiment --- each shown on a vertical cross-section of the sample
    (left) and on the fault plane (right); the subcaptions give each stage's
    equivalent fault slip $s$. All panels share the same colour scale. The
    equivalent figure for all 23 stages is given in the Supporting
    Information.}
    \label{fig:acustic_emission_density_selected_stages}
\end{figure}

The AE event density field, which is independent of the inversion, describes
where damage accumulates.
Figure~\ref{fig:acustic_emission_density_selected_stages} shows the field at
eight representative stages, on a vertical cross-section through the sample and
on the fault plane itself (the corresponding figure for all 23 stages is given
in the Supporting Information).

In the early stages two damage zones propagate from the top and bottom notches
and remain disconnected. These zones grow within the fault volume and become
connected during stages 10--13 ($s=0.42$--$0.52$~mm), in the final weakening phase
\protect\cite{Aben2024}. In the later stages the region of high AE density
continues to expand.

\begin{figure}
    \centering
    \includegraphics[width=\linewidth]{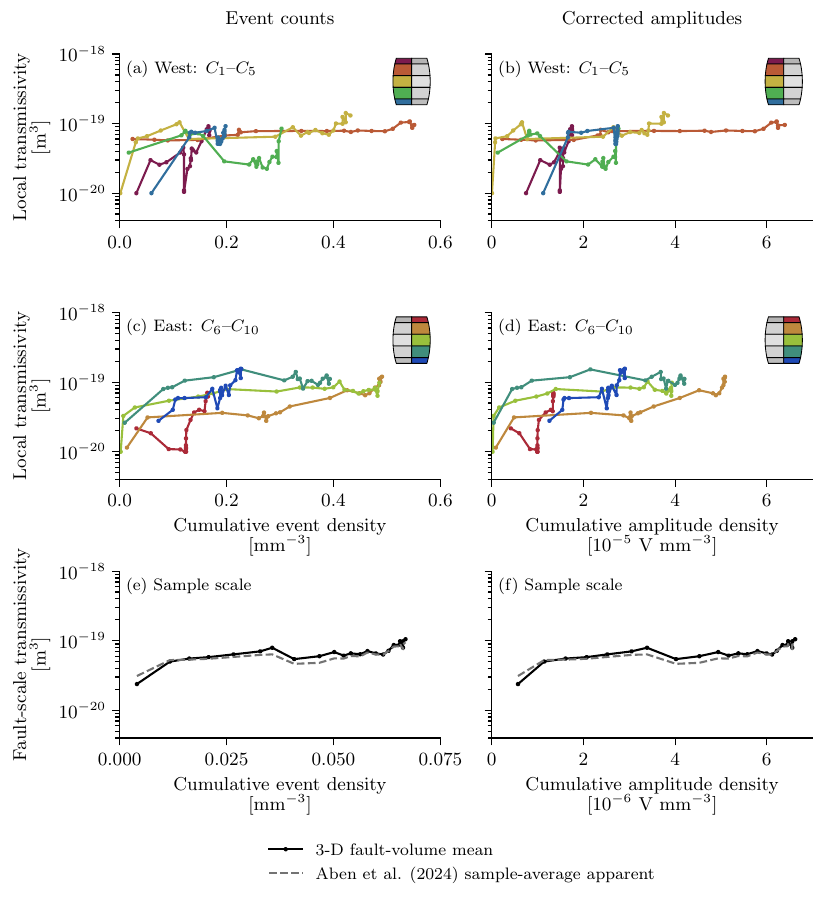}
    \caption{Transmissivity against cumulative AE event density (left) and
    corrected-amplitude density, accumulated from $A_e=10^{M_{\mathrm{AE},e}}$
    (right), using the same events in both columns. (a,b) Western cells
    $C_1$--$C_5$; (c,d) eastern cells $C_6$--$C_{10}$. The insets identify
    cells in the colours of Figure~\ref{fig:inverse_problem_unknowns}.
    Local densities use the same Gaussian spatial averaging and cell volumes.
    (e,f) Fault-volume-weighted mean of the ten 3-D transmissivities (solid)
    and sample-average apparent transmissivity of \protect\citeA{Aben2024}
    (dashed), against catalogue totals divided by sample volume.
    Points mark hydraulic measurements and lines follow their chronological
    order. Amplitudes are in volts; note the different
    horizontal scales for local and sample-wide densities.}

    \label{fig:permeabilities_vs_ae}
\end{figure}

Figure~\ref{fig:permeabilities_vs_ae} shows the local transmissivity of each
cell against cumulative AE event density and corrected-amplitude density,
together with the
fault-volume-weighted mean transmissivity and the sample-average apparent
transmissivity of \protect\citeA{Aben2024} against the corresponding
sample-wide quantities. Both columns use the same events with amplitudes.
The sample-average apparent value grows by about a factor of three over the
test, consistent with the post-peak increase in average permeability measured
by \protect\citeA{Aben2024}. The individual cells follow no common path: at a given local AE event or
corrected-amplitude density the local transmissivity differs between cells,
and several cells show a transmissivity that decreases while their AE
activity continues to accumulate. A shared monotonic relationship is not
apparent across all ten cells.

The additional comparisons separate cumulative growth from incremental
hydraulic change (Supporting Information Tables~S4 and S5). Cumulative event
count and cumulative corrected amplitude each have a Spearman correlation of
0.81 with the fault-volume-weighted mean transmissivity across the 23
measurements. Across the 22 successive intervals, the
correlations between local AE-density increments and
$\log_{10}(T_j/T_{j-1})$ range from $-0.34$ to 0.42 among cells; excluding
the three intervals spanning overnight holds gives $-0.17$ to 0.62
($n=19$). The 120~s aggregate
catalogue rate has a correlation of $-0.16$ with changes in the mean
transmissivity. These descriptive comparisons distinguish aggregate
co-evolution from the variable local associations between recent AE activity
and hydraulic change. Including event size retains this spatial variability:
on the amplitude-matched subset, correlations between amplitude-weighted
local activity and $\log_{10}(T_j/T_{j-1})$ range from $-0.25$ to 0.38,
compared with $-0.34$ to 0.42 for unweighted activity on the same events
(Supporting Information Table~S5).

\section{Discussion}

\subsection{How far can the recovered structure be trusted?}
The synthetic ensembles of Section~\ref{sec:verification} recover the two
tested structures with median mean errors of 7--9 \% at the gauge-accuracy
scale. The fault-volume-weighted mean also varies little across initial guesses
(median 2.6 \%) and the tested off-fault permeabilities (median 7.4 \%;
Supporting Information Text S3), supporting the fault-scale history. Ten
parameters are constrained by seven observations, so individual cell values are
not unique. Cells $C_{1}$--$C_{5}$ are constrained by two retained gauges and
cells $C_{6}$--$C_{10}$ by four; the two sides of the fault are therefore not
imaged equally well.

The thickness tests of Section~\ref{sec:verification} have two practical
consequences. Absolute permeabilities are conditional on the assumed thickness,
whereas the corresponding transmissivities are much steadier: the median
absolute cellwise change is at most about 16 \% over the 5--15~mm range. This
supports reporting the effective transmissivity $k_ft$ that
\protect\citeA{Aben2024} adopted in their first-order estimates.

\subsection{Heterogeneity and fault structure}
The fault transmissivity recovered from the inverse method is qualitatively
consistent with the earlier, simpler estimates of \protect\citeA{Aben2024}. The
quantitative differences are at most of the order of a factor 2, and are due to
the difference in flow patterns that are fully accounted for in our new
approach. The conclusions of Aben et al. regarding the development of
permeability heterogeneity during fault growth are therefore reinforced by our
new results: permeability development appears to be dominantly controlled by
the connectivity of fault subbranches, a process that occurs episodically
during fault formation and growth. The transmissivity heterogeneity is very
strong at early stages, with variations of around one order of magnitude
between different cells. With ongoing fault connection and slip, there is a
clear trend towards homogenisation of transport properties, with only smaller
variations of the order of 50\% persisting once residual friction and fault
completion is achieved. The early transmissivity increase accompanies the
propagation of damage away from the two notches. A later increase occurs after
the AE-defined damage zones become connected over $s=0.42$--$0.52$ mm. Opening,
closure and connection of individual fault branches can produce both increases
and decreases in flow capacity. This is a mechanistic interpretation rather
than a direct image of the conducting apertures, but it explains why
transmissivity changes episodically while cumulative damage continues to grow.

The heterogeneity late in the experiment, and in particular the high
transmissivity of $C_{10}$ near the downstream notch, persists across initial
guesses (Figure~\ref{fig:sensitivity_optimised_ci}), although the exact cell
ranking can vary. A plausible explanation lies in the local geometry of the
fault: Near this notch the fault may include a more tensional strand, which
would open rather than shear and would therefore be more transmissive than the
rest of the plane.

One important additional source of transmissivity variation, unanticipated
while designing the original experiment, is the unloading (down to 153~MPa
differential stress), 14.5-hour hold, and re-loading between stages 7 and 8.
This sequence was used to momentarily pause the experiment overnight, and
repeated again between stages 15 and 16, and 21 and 22. During the first hold
between stages 7 and 8, most cells experience a decrease in transmissivity
across this sequence that is not reversed upon re-loading. Such a change was
not observed during subsequent holds, conducted after completion of the fault
across the entire sample. The transmissivity decrease during the first hold was
already noted in our earlier estimates using uniform flow assumption, but was
not found as quantitatively strong. In particular, cells $C_{1}$ and $C_{4}$
show an approximately threefold drop, and only recover slowly as slip proceeds.
In principle, such hysteresis can be explained by local back-slip and fracture
closure during unloading. Local back-slip on the fault is driven by elastic
restoring forces generated by the surrounding material (e.g.,
\protect\citeNP{David2012}): these forces can exist only if the fault is
contained (at least partially) within the specimen, i.e., before it reaches the
free surfaces of the cylinder. The lack of significant transmissivity
hysteresis during unloading/hold/re-loading sequences when the fault has become
through-going is thus consistent with the role of back-slip at early stages. To
further confirm this explanation, spatially resolved strain or slip
measurements would be required.

The large local transmissivity decrease inferred after unloading between stages
7 and 8, when the fault is still partially contained within the sample, is at
odds with the permeability hysteresis typically observed before fault formation
in granite (\protect\citeNP{Zoback1975}; \protect\citeNP{Mitchell2008}). In the
pre-peak stage, permeability tends to remain high, near its previous achieved
maximum, as differential stress is decreased: this hysteresis likely originates
from the frictional forces along preexisting sliding defects that prevent
backslip and keep tensile microcracks open (e.g.,
\protect\citeNP{ZobackByerlee1975Cyclic}). By contrast, in our partially
faulted sample, fault zone transmissivity decreases more readily upon
unloading, and remains low even after re-loading. The relative ease of fault
back-slip and fracture closure could be due to the more subvertical orientation
of fault strands near the notches
(Figure~\ref{fig:acustic_emission_density_selected_stages}b), as well as the
overall longer fracture length that makes the fracture system more compliant.
However, the origin of the persistence of the transmissivity decrease is more
difficult to establish. One possibility is that some time-dependent compaction
or restrengthening occurred during the 14.5~h hold period, which prevented
direct reopening of the flow path upon re-loading.

Taken together, the inversion results show that transmissivity is impacted not
only by fault growth, connectivity and slip, but also by deformation history
and stress state. The heterogeneity and episodicity of transmissivity evolution
are therefore not only driven by purely structural processes (the fault
formation process itself), but also by other ambient conditions (here, stress
and stress history).

\subsection{Acoustic emissions and local transmissivity}
The AE maps locate the growth and linkage of the damage zones, while the
hydraulic inversion quantifies how their flow capacity evolves. Sample-average
apparent transmissivity increases overall as cumulative AE activity grows,
which is the co-evolution visible to sample-scale measurements. Locally,
however, cells with comparable cumulative AE event or corrected-amplitude
densities differ in transmissivity by up to about an order of magnitude, and
some cells become less transmissive while their AE density continues to
increase: there is no consistent correlation between AE activity (number or
accumulated corrected amplitude) and transmissivity change at the cm-scale.
This observation in a dilatant, brittle granite is consistent with recent work
on compactant porous sandstone, which shows that AE number density tracks
wave-speed changes more consistently than local permeability change, the latter
depending strongly on the geometry and connection of compaction bands
(\protect\citeNP{BrantutBaud2025}).

During brittle deformation of low porosity crystalline rocks, AE activity is
typically linked to tensile or shear microcracking, which are linked to
dilatancy and are therefore expected to promote permeability increases: this is
indeed what happens in the pre-peak stage for monotonically loaded samples
(\protect\citeNP{PatersonWong2005}; Section 5.8). In the post-peak stage,
strain localisation in the form of a shear fault changes dramatically how
permeability is generated and how this generation is manifested as AE activity.
Firstly, fault slip is initially linked to local dilation only if the fault
zone material is overconsolidated: this is necessarily true at the onset of
slip, but dilation might saturate quickly, and even give way to compaction as
fine-grained fault gouge is formed. Slip on preexisting fault or fault gouge is
indeed systematically associated with dramatic permeability reduction (e.g.,
\protect\citeNP{ZhangTullis1998}; \protect\citeNP{RutterMecklenburgh2018}),
while producing consistent AE activity (\protect\citeNP{Mair2007}). In
addition, the process of fine-grained fault gouge formation from intact rock
may require a remarkably limited amount of slip (e.g.,
\protect\citeNP{Incel2025}). Secondly, the post-peak stage is marked by a
gradual, macroscopic differential stress drop, and fault expansion is expected
to generate local, transient stress heterogeneities. As discussed in the
previous section, there is considerable hysteresis in permeability during
unload-reload cycles, whereas AE activity is nearly nonexistent there
(consistently with the widely reported Kaiser effect, see
\protect\citeA{Lavrov2003}). The role of stress state could locally dominate
the permeability response, compared to the role of fault slip. Thirdly, the
increase in connectivity of fault branches is a discontinuous process, not
necessarily associated with detectable variations in local AE activity.
Finally, the entire faulting process is associated to substantial aseismic
(here, acoustically silent) deformation that impacts local permeability in ways
that cannot be detected by AE activity monitoring (e.g.,
\protect\citeNP{Cartwright-Taylor2022}).

Our results provide useful constraints for the interpretation of
reservoir-scale seismicity patterns. Seismicity nearly systematically occurs in
response to fluid injection at depth, but it is not always clear if induced
seismicity impacts permanently the fluid transport properties of the reservoir
rock. A link between seismicity and permanent permeability change was
established in the case of the Basel injection experiment
(\protect\citeNP{Haring2008}), where a sustained 400-fold increase in fracture
transmissivity was inferred following injection that led to a seismic sequence
culminating with a $M_L\,3.4$ earthquake. Correlations between the
spatio-temporal and magnitude evolution of induced microseismicity and
injectivity (a proxy for permeability) in enhanced geothermal reservoirs have
been interpreted as a direct causality between seismic slip, fault aperture
change, and permeability change (e.g., \protect\citeNP{RothertBaisch2010};
\protect\citeNP{Fang2018}; \protect\citeNP{Yu2024}; \protect\citeNP{Yu2026}). A
contrasting case study is the experiment of Espoo, Finland, where fluid was
injected at around 6.1 km depth in an intact granite body
(\protect\citeNP{Kwiatek2019}). The injection produced abundant seismicity,
with moment magnitudes up to 1.9, and injectivity (a proxy for permeability)
transiently increased during the injection procedure. However, injectivity
decreased back to its initial background level after injection stopped
(\protect\citeNP{Kukkonen2023}). Our laboratory data and analysis suggest that
seismicity alone is not a universal predictor of hydraulic transmissivity, and
that the effect of fault slip (seismic or aseismic) may not be dominant over
the effect of stress state or stress history: in the case of Espoo, the fault
geometry and stress state were such that the fluid pressure decrease after
injection led to fracture closure, which is analogous to the local
transmissivity decrease observed in response to unloading in our experiment.

The analogy between AE activity in the laboratory and seismicity in the field
is limited: a proportion of AE events are inherently at the grain scale (e.g.,
grain crushing or pore collapse), and only shear events should be considered
when extrapolating to larger scales (\protect\citeNP{Blanke2021}). Despite this
limitation, our data suggest that a direct link between seismic activity or
moment and transmissivity increase is not universal, and may only exist subject
to conditions in terms of stress state, rock type, and preexisting fault
structures. An extreme example where AE activity is linked to a decrease, as
opposed to an increase, in permeability, is the case of porous sandstone
undergoing compaction (e.g., \protect\citeNP{Fortin2006};
\protect\citeNP{BrantutBaud2025}). Even then, the connectivity of compacted
regions showing high AE activity plays a substantial role, and not simply the
cumulative AE activity itself.

\section{Conclusions}
We developed a three-dimensional finite-element inverse method to estimate the
local hydraulic transmissivity along a fault in Westerly granite, from the
measured fluid flow rate and six internal pore pressure measurements obtained
at 23 stages of a quasi-static shear failure test. We find that the local
transmissivity reaches approximately $1.6\times10^{-19}$ m$^3$ (16 times the
bulk contribution $K_0t$). Its evolution is episodic, with early and later
increases interspersed with decreases, consistent with changes in opening and
connectivity. The fault is heterogeneous, with a factor of 2.3 difference
between local transmissivities at the final stage, while the sample-average
apparent transmissivity grows about threefold over the whole test. Our full 3D
inversion agrees in scale with earlier results that neglected flow
redistribution, but we detect stronger local variability at early stages of
rupture propagation, with transmissivity ratios of up to one order of magnitude
between cells. Across cells, the median absolute change in recovered
transmissivity is at most about 16 \% for $\pm50$ \% changes in assumed
fault-zone thickness, and the equivalent permeabilities scale roughly inversely
with it: we therefore only constrain hydraulic transmissivity and not
permeability alone.

Partial unloading at a stage where the shear fault is not yet through-going
leads to substantial local changes in transmissivity that are not fully
recovered during re-loading. This indicates irreversible fracture closure
possibly driven by local backslip and compaction.

A detailed comparison between hydraulic transmissivity evolution and AE
activity shows a lack of consistent local correlation between either cumulative
AE number density or accumulated corrected amplitude and transmissivity
increase, even though such a correlation exists at the sample scale. In
particular, the local transmissivity evolution may not be monotonic with
increasing AE activity, indicating the role of gouge formation and compaction
during the overall shear fracturing process.

Our results are consistent with field-scale relationships between microseismic
activity and permeability evolution: AE activity at reservoir scale indicates a
trend towards permeability increase, but this increase may not be permanent and
might be highly variable locally depending on fault branch connectivity and
stress state.

\section*{Open Research}
The experimental data underlying this study are those of \protect\citeA{Aben2024},
available at \protect\citeA{AbenBrantutDataset2024}. The processed
stress--displacement series used to regenerate Figure~\ref{fig:experimental_overview}c,
the analysis scripts, the derived local apparent-transmissivity comparison table,
the supplied distance-corrected AE amplitude catalogue, and the count-based
and relative-amplitude AE metric tables are included in the accompanying
manuscript repository.

\acknowledgments
Neil Hughes provided technical support on the triaxial apparatus. We
acknowledge funding from the UK Natural Environment Research Council (grant
NE/S000852/1) and the European Research Council under the European Union's
Horizon 2020 research and innovation programme (project RockDEaF, grant
agreement 804685), and from the European Research Council (project RockDeath,
grant agreement 101088963). Nicolas Brantut was also supported by the Philip
Leverhulme Prize from the Leverhulme Trust.

\end{document}


\title{Supporting Information for ``Heterogeneous and episodic permeability
evolution and correlation with acoustic emission activity during faulting in
granite''}

\authors{Ado Farsi\affil{1,2,3}, Franciscus M. Aben\affil{4}, and
Nicolas Brantut\affil{1,5}}

\affiliation{1}{Department of Earth Sciences, University College London, London, UK}
\affiliation{2}{Department of Mathematics, Imperial College London, London, UK}
\affiliation{3}{Tanuki Technologies, London, UK}
\affiliation{4}{TNO Applied Geosciences, Utrecht, The Netherlands}
\affiliation{5}{Geomechanics and Scientific Drilling, GFZ Helmholtz Centre for Geosciences, Potsdam, Germany}

\section*{Contents of this file}
\begin{enumerate}
\item Text S1 and Table S1: local comparison with Aben et al. (2024)
\item Text S2 and Table S2: pre-deformation spatial inversion
\item Text S3 and Table S3: sensitivity to the off-fault permeability
\item Figure S1: acoustic-emission event density at all measured slip values
\item Text S4 and Table S4: count-based AE metrics and hydraulic changes
\item Text S5 and Table S5: relative AE amplitudes and hydraulic changes
\end{enumerate}

This Supporting Information gives the local 3-D/apparent comparison, the
pre-deformation reference and off-fault-permeability sensitivity, and
acoustic-emission (AE) maps for all 23 flow measurements. Figure~S1 extends
the main-text figure at eight slip values using the same colour scale.
Text S4 adds comparisons of cumulative AE counts, recent accumulation and
short-window catalogue rates with the hydraulic results. Text S5 extends
the comparison to relative event sizes.

\enlargethispage{2\baselineskip}
\section*{Text S1. Local apparent-transmissivity reconstruction}

For each of the 23 post-peak steady-flow measurements, we reproduced Equations~5--6
of \protect\citeA{Aben2024}. In region $i$ their one-dimensional approximation is
\begin{equation}
q=\langle\ell\rangle_i
  \frac{\langle k_f t\rangle_i}{\mu}
  \frac{\Delta p_i}{\Delta X_i},
\label{eq:si_aben_local_transmissivity}
\end{equation}
where $q$ is the measured flow rate, $\mu=10^{-3}$~Pa~s,
$\langle\ell\rangle_i$ is the mean width of the elliptical fault plane over
the region, $\Delta p_i$ is the measured local pressure drop, and $\Delta X_i$
is the down-dip distance between its bounding measurements or notch tip. The
measured $q$ is assigned to every region because local flow rates were not
measured. The resulting $\langle k_f t\rangle_i$ is therefore an apparent
estimate under a one-dimensional allocation of the measured total flow to
every serial region; it neglects
off-fault and cross-fault flow. \protect\citeA{Aben2024} interpreted such
estimates as lower bounds on local transmissivity \emph{changes}, because a
locally more permeable region could carry more than its assumed share of the
flow, while also cautioning that three-dimensional effects prevent a general
conversion to absolute local values.

The pressure locations divide each side into five regions. Their boundaries
and those of the three-dimensional cells differ by at most 0.901~mm, allowing
a region-by-region comparison without interpolation. On the Western side,
$P_1$ was unavailable throughout the experiment, so the upstream-to-$P_2$
estimate is assigned to both upstream regions exactly as in Figure~11 of
\protect\citeA{Aben2024}; those two values are shared rather than independent.
Gauge $P_4$ drifted substantially after the $s=0.60$~mm measurement. The two
affected downstream Western estimates at each of the eight subsequent pauses
($s=0.56$--$1.41$~mm) remain in Table~S1 (asterisks) to reproduce the
published calculation but are excluded from the ratio summary. No other
rows are excluded.

\clearpage
Table~\ref{tab:si_aben_local_comparison} summarises comparisons matched by
measurement, side and down-dip region. Excluding the 16 late
Western comparisons affected by $P_4$ drift, the 214 cell--measurement ratios
of 3-D to apparent transmissivity have a median of 0.81 and a descriptive
5th--95th percentile range of 0.44--1.92. The estimates overlap in scale;
regional differences reflect the different flow allocations. Ratios below unity do not
contradict the qualified lower-bound interpretation of \protect\citeA{Aben2024},
which concerns local \emph{changes}, not an absolute ordering between models.
These pooled ratios describe one experiment; they are not independent
statistical observations or uncertainty intervals.

\begin{table}[htbp]
\caption{Local comparison at the 23 measured equivalent-slip values. ``Aben average'' is the
sample-average apparent transmissivity; ``Aben local'' and ``3-D local'' give
the ranges across ten regions/cells. The final column is the median
[minimum--maximum] cellwise ratio after excluding the explicitly
$P_4$-drift-affected rows. Transmissivities are in $10^{-20}$~m$^3$ and $s$
is equivalent fault slip. Stage numbers are retained as chronological data
indices. An asterisk marks measurements whose reported Aben local range
includes the two retained $P_4$-affected Western estimates.}
\label{tab:si_aben_local_comparison}
\centering
\scriptsize
\begin{tabular}{rrrrrr}
\toprule
Stage & $s$ [mm] & Aben average & Aben local & 3-D local &
3-D/Aben \\
\midrule
1 & 0.06 & 3.11 & 2.27--5.47 & 1.00--6.00 & 0.46 [0.28--1.67] \\
2 & 0.15 & 5.27 & 2.55--13.29 & 1.84--7.96 & 0.73 [0.46--1.41] \\
3 & 0.17 & 5.35 & 2.07--12.87 & 1.09--8.30 & 0.71 [0.45--1.50] \\
4 & 0.19 & 5.50 & 2.11--12.79 & 1.08--8.44 & 0.73 [0.44--1.50] \\
5 & 0.24 & 5.85 & 2.27--15.28 & 1.00--10.52 & 0.84 [0.40--1.57] \\
6 & 0.31 & 6.26 & 2.49--16.91 & 1.24--11.79 & 0.90 [0.42--1.50] \\
7 & 0.35 & 6.37 & 2.24--19.73 & 1.00--15.26 & 0.80 [0.43--1.84] \\
8 & 0.30 & 4.65 & 1.79--13.47 & 1.10--10.60 & 0.64 [0.32--2.22] \\
9 & 0.37 & 4.82 & 1.70--13.77 & 1.00--11.95 & 0.65 [0.31--2.32] \\
10 & 0.42 & 5.58 & 2.04--16.03 & 1.15--14.02 & 0.66 [0.49--1.95] \\
11 & 0.46 & 5.54 & 2.23--12.32 & 1.56--11.05 & 0.71 [0.48--1.82] \\
12 & 0.50 & 6.03 & 2.53--12.76 & 2.05--11.26 & 0.78 [0.54--1.85] \\
13 & 0.52 & 5.96 & 2.74--13.12 & 2.61--10.20 & 0.83 [0.51--1.69] \\
14 & 0.55 & 6.77 & 3.23--13.69 & 3.19--10.34 & 0.82 [0.59--1.90] \\
15 & 0.60 & 6.36 & 3.31--13.26 & 2.34--10.45 & 0.85 [0.46--1.71] \\
16 & 0.56 & 6.32 & 3.21--12.75$^{*}$ & 2.24--9.14 & 0.85 [0.54--1.98] \\
17 & 0.64 & 7.10 & 3.98--13.71$^{*}$ & 2.80--10.68 & 0.88 [0.58--2.08] \\
18 & 0.72 & 8.25 & 4.74--16.04$^{*}$ & 3.35--14.85 & 0.99 [0.63--2.34] \\
19 & 0.81 & 8.32 & 4.75--16.09$^{*}$ & 3.30--13.79 & 1.00 [0.62--2.27] \\
20 & 0.90 & 8.95 & 5.01--19.22$^{*}$ & 4.03--14.94 & 1.01 [0.65--2.30] \\
21 & 1.05 & 7.54 & 4.15--17.35$^{*}$ & 2.73--11.43 & 0.92 [0.60--2.12] \\
22 & 1.22 & 8.88 & 4.75--21.05$^{*}$ & 5.57--15.60 & 0.95 [0.71--2.23] \\
23 & 1.41 & 9.49 & 4.81--20.31$^{*}$ & 6.85--15.48 & 0.94 [0.73--2.13] \\%
\bottomrule
\end{tabular}
\end{table}
\clearpage

\section*{Text S2. Pre-deformation spatial inversion}

The WGMS4 reference flow test was acquired at a differential stress of
$-5.88$~MPa, before deformation and fault growth. First, we calibrated a
homogeneous permeability to its measured flow of
$2.072\times10^{-12}$~m$^3$~s$^{-1}$ using the same geometry, conductive
notches and boundary conditions as the post-peak calculations. This gives
$K_\mathrm{pre}=2.092\times10^{-20}$~m$^2$. Equivalently, the apparent
permeability from Equations~2--3 of \protect\citeA{Aben2024} is
$1.403\times10^{-20}$~m$^2$ and multiplication by the independently
reproduced geometry correction $C=1.491$ gives the same value. The length in
this calculation is the 42.453~mm axial separation of the inner borehole
tips. The homogeneous-model pressure RMSE is 0.228~MPa (Table~S2a).

We then inverted the measured flow and the same six internal pressures for
ten independent permeabilities in the prospective fault cells, holding the
surrounding rock at $K_\mathrm{pre}$. The mesh, finite elements and relative
least-squares objective are those of the post-peak inversions. The initial
cell permeabilities are all $K_\mathrm{pre}$, with bounds
$0.1K_\mathrm{pre}\leq k_f^i\leq10K_\mathrm{pre}$ to allow variation on
either side of the intact-rock reference; no homogeneity penalty is applied.

The recovered local permeabilities range from
$4.98\times10^{-21}$ to $3.72\times10^{-20}$~m$^2$ (Table~S2b). All remain
at the intact-rock scale, with spatial variation rather than a uniform
field. The pressure RMSE falls to $4.72\times10^{-5}$~MPa and the relative
flow residual is $-0.00050$~\%; no cell reaches either bound.
Repeating the inversion from alternating $0.8K_\mathrm{pre}$ and
$1.2K_\mathrm{pre}$ starting values gives a median absolute cellwise change
of 5.0~\%, with a similarly close fit and no active bounds.

\begin{table}[p]
\caption{Pre-deformation WGMS4 results. (a) Measured pore pressures,
    homogeneous-model predictions and spatial-inversion predictions.
    Residual is spatial prediction minus measurement. (b) Recovered local
    permeabilities for the homogeneous and perturbed starting points, in
    $10^{-20}$~m$^2$. Gauges $P_1$ and $P_4$ are omitted to retain the same
    six-gauge observation set as the post-peak inversions.}
\label{tab:si_predeformation_pressures}
\centering
\small
(a) Pore pressure comparison\\[4pt]
\begin{tabular}{rrrrr}
\toprule
Gauge & Measured & Homogeneous & Spatial & Residual \\
& [MPa] & [MPa] & [MPa] & [MPa] \\
\midrule
$P_2$ & 3.210 & 2.923 & 3.210 & $-0.00001$ \\
$P_3$ & 2.515 & 2.065 & 2.515 & $-0.00001$ \\
$P_5$ & 3.716 & 3.824 & 3.716 & $+0.00007$ \\
$P_6$ & 2.910 & 2.923 & 2.910 & $-0.00009$ \\
$P_7$ & 2.178 & 2.065 & 2.178 & $+0.00000$ \\
$P_8$ & 1.220 & 1.165 & 1.220 & $+0.00001$ \\
\bottomrule
\end{tabular}

\vspace{12pt}
(b) Local permeability\\[4pt]
\begin{tabular}{rrr}
\toprule
Cell & Homogeneous start & Perturbed start \\
\midrule
$C_{1}$ & 3.028 & 3.226 \\
$C_{2}$ & 3.718 & 3.498 \\
$C_{3}$ & 2.272 & 2.180 \\
$C_{4}$ & 0.498 & 0.408 \\
$C_{5}$ & 1.417 & 2.162 \\
$C_{6}$ & 1.337 & 1.314 \\
$C_{7}$ & 2.293 & 2.272 \\
$C_{8}$ & 2.976 & 2.932 \\
$C_{9}$ & 2.418 & 2.336 \\
$C_{10}$ & 2.486 & 2.331 \\
\bottomrule
\end{tabular}
\end{table}
\clearpage

\section*{Text S3. Sensitivity to the off-fault permeability}

The primary post-peak inversions use the rounded, prescribed off-fault
permeability $K_0=10^{-18}$~m$^2$. We also repeated the complete sequential
23-measurement inversion with $K_0=7.5\times10^{-19}$~m$^2$, the measured
pre-fault-growth \emph{sample-average apparent} permeability used here as an
alternative off-fault assumption. The inversion for the first post-peak
measurement ($s=0.06$~mm) is initialised at $C_i=1$; each later inversion is
initialised from the preceding solution exactly as in the primary chain. This is a
new inversion in which both the off-fault flow and the dimensional
fault-zone permeability $k_f^i=C_i K_0$ change; it is not an algebraic
rescaling of the primary coefficients. Table~S3 reports three representative
slip values. The comparison tests the prescribed background permeability.

The lower $K_0$ removes the nominal chain's lower-bound limitation at
$s=0.06$~mm: none of the 230 alternative coefficients reaches the lower bound
and the objective at $s=0.06$~mm falls from $5.80\times10^{-3}$ to
$2.11\times10^{-10}$. The
fault-volume-weighted mean transmissivity differs from the nominal chain by a
median of 7.4~\% and at most 10.9~\% across measurements. Local differences are less
uniform: the median absolute cellwise difference is 9.3~\% over the complete
cell--measurement set, and is 4.3~\% at $s=0.60$~mm and 7.6~\% at
$s=1.41$~mm. These results support a similar fault-scale history for both prescribed
off-fault permeabilities.

\begin{table}[h]
\caption{Sensitivity to $K_0$. Transmissivity ranges are across cells
$C_1$--$C_{10}$ and are in $10^{-20}$~m$^3$. ``Median change'' is the median
absolute cellwise difference relative to the primary $K_0=10^{-18}$~m$^2$
chain. $J$ is the seven-observation objective defined in the main text.}
\label{tab:si_k0_sensitivity}
\centering
\small
\begin{tabular}{rccrrr}
\toprule
& \multicolumn{2}{c}{Local transmissivity range} & &
\multicolumn{2}{c}{$J$} \\
\cmidrule(lr){2-3}\cmidrule(lr){5-6}
$s$ [mm] & $K_0=10^{-18}$ & $K_0=7.5\times10^{-19}$ &
Median change [\%] & $10^{-18}$ & $7.5\times10^{-19}$ \\
\midrule
0.06 & 1.00--6.00 & 0.88--3.35 & 33.4 & 5.80e-03 & 2.11e-10 \\
0.60 & 2.34--10.45 & 3.50--10.61 & 4.3 & 6.05e-10 & 4.31e-10 \\
1.41 & 6.85--15.48 & 7.20--16.83 & 7.6 & 2.31e-09 & 3.16e-10 \\%
\bottomrule
\end{tabular}
\end{table}
\clearpage

\begin{figure}[p]
\centering
\includegraphics[width=\linewidth,height=0.82\textheight,keepaspectratio]{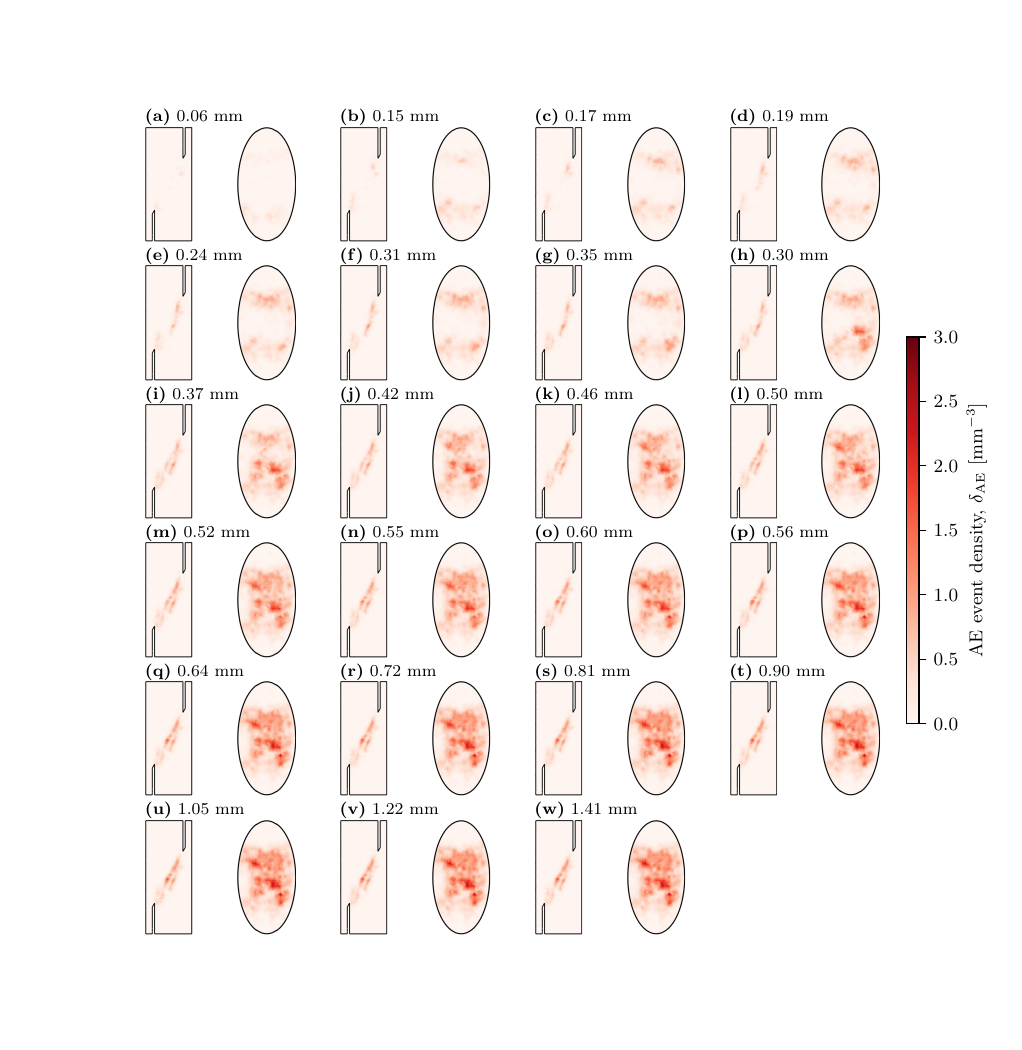}
\caption{Cumulative AE event-density field at all 23 steady-flow
measurements: (a)--(w) are ordered chronologically. At each slip value the field is shown on a
vertical cross-section of the sample (left) and on the fault plane (right),
with the equivalent fault slip above each pair. All panels share the colour
scale of the corresponding main-text figure.}
\label{fig:si_acustic_emission_density_all_stages}
\end{figure}
\clearpage

\section*{Text S4. Count-based AE metrics and hydraulic changes}

The located-event catalogue contains 9,932 rows with time and three spatial
coordinates. We retain the audited stage boundaries used for the AE fields:
9,913 events precede the final boundary and 19 follow it. These count-based
comparisons give every event equal weight. The separate amplitude catalogue
is analysed in Text S5.

Let $N_j$ be the cumulative located-event count at measurement $j$ and
$\delta_{i,j}$ the existing Gaussian AE density averaged over cell $i$.
For measurements $j=2,\ldots,23$, define
\begin{equation}
\Delta N_j=N_j-N_{j-1},\qquad
\Delta\delta_{i,j}=\delta_{i,j}-\delta_{i,j-1},\qquad
D_{i,j}=\log_{10}\!\left(\frac{T_{i,j}}{T_{i,j-1}}\right),
\end{equation}
where $T_{i,j}=\langle k_f t\rangle_{i,j}$. These 22 post-peak intervals
compare recent event accumulation with signed hydraulic changes. We also use
the fault-volume-weighted mean
\[
\overline{T}_j=\frac{\sum_i V_iT_{i,j}}{\sum_i V_i}.
\]
Its change is $\log_{10}(\overline{T}_j/\overline{T}_{j-1})$, not the
average of the local logarithmic changes.

To distinguish event rate from cumulative count, we calculate
\begin{equation}
\lambda_{w,j}=\frac{1}{w}\sum_e
\mathbf{1}\{t_j-w<t_e\leq t_j\},
\end{equation}
where $t_e$ is event time, $t_j$ is the last located event before the audited
pause boundary, and $w=120$~s, with 60 and 300~s sensitivity windows.
These are sample-wide, endpoint-conditioned catalogue rates, not average
rates over a complete loading stage. Windows can include quiet time, and
the catalogue does not provide a time-dependent detection-completeness
correction. The rate comparison is therefore descriptive and does not
replicate a field stimulation model or its event-size features.

Table~\ref{tab:si_ae_count_metrics} reports Spearman rank correlations, using
average ranks for ties. Cumulative quantities describe the shared evolution
over 23 measurements; increments describe 22 consecutive changes. Local
correlations are calculated separately for each cell. The third local column
excludes intervals ending at stages 8, 16 and 22, the three
intervals spanning overnight holds. We do not treat the cells or successive measurements as
independent replicates, report significance tests, or infer predictive skill
from these in-sample correlations.

Using 60 and 300~s windows gives correlations of $-0.17$ and $-0.05$,
respectively, between catalogue rate and changes in mean transmissivity,
compared with $-0.16$ at 120~s. Repeating the incremental local comparison
with the existing alternative-initialisation and alternative-background
inversions gives cellwise correlation ranges of $-0.33$ to $0.48$ and
$-0.28$ to $0.42$, respectively. The aggregate cumulative-count correlations
remain 0.81 and 0.80. These checks retain the distinction between overall
co-evolution and variable local incremental associations. The accompanying
analysis data retain every stage and cell value, all three windows, and an
additional check excluding the first post-peak interval together with the
unload--reload transitions.

\begin{table}[htbp]
\caption{Descriptive Spearman correlations for the count-based AE comparisons.
(a) Aggregate metrics, where $n$ is the number of paired measurements or
intervals. (b) Local metrics for all ten cells: cumulative density against
$T$ ($n=23$ per cell), and density increments against $D$ using all intervals
($n=22$) or excluding the three intervals spanning overnight holds ($n=19$).
The coefficients describe this experiment and are not tests of a universal
relation or measures of prediction on independent data.}
\label{tab:si_ae_count_metrics}
\centering
\small
(a) Aggregate comparisons\\[4pt]
\begin{tabular}{llrr}
\toprule
AE metric & Hydraulic quantity & $n$ & $r_s$ \\
\midrule
Cumulative count $N_j$ & $\overline{T}_j$ & 23 & 0.81 \\
Count increment $\Delta N_j$ & $\Delta\log_{10}\overline{T}_j$ & 22 & -0.01 \\
Catalogue rate $\lambda_{120,j}$ & $\overline{T}_j$ & 23 & -0.50 \\
Catalogue rate $\lambda_{120,j}$ & $\Delta\log_{10}\overline{T}_j$ & 22 & -0.16 \\%
\bottomrule
\end{tabular}

\vspace{12pt}
(b) Local comparisons\\[4pt]
\begin{tabular}{lrrr}
\toprule
Cell & Cumulative & Increments & Excluding unload--reload \\
\midrule
$C_{1}$ & 0.77 & 0.42 & 0.47 \\
$C_{2}$ & 0.88 & -0.12 & -0.15 \\
$C_{3}$ & 0.75 & 0.00 & 0.16 \\
$C_{4}$ & -0.28 & -0.34 & -0.14 \\
$C_{5}$ & 0.21 & 0.01 & 0.26 \\
$C_{6}$ & 0.84 & -0.22 & -0.17 \\
$C_{7}$ & 0.93 & 0.20 & 0.22 \\
$C_{8}$ & 0.48 & -0.03 & 0.10 \\
$C_{9}$ & 0.31 & 0.24 & 0.62 \\
$C_{10}$ & 0.93 & 0.02 & 0.21 \\%
\bottomrule
\end{tabular}
\end{table}
\clearpage

\section*{Text S5. Relative AE amplitudes and hydraulic changes}

The supplied catalogue contains 8,415 positive, finite amplitudes, corrected
for source--receiver geometrical distance following \protect\citeA{Zang2000}.
Although its final column is labelled ``Mag'', it contains linear amplitudes
$A_e$ in volts; we take their logarithm once to obtain
$M_{\mathrm{AE},e}=\log_{10}A_e$. This proxy describes relative event size
within the experiment; neither $A_e$ nor $M_{\mathrm{AE},e}$ is converted to
seismic moment or energy.

We match all 8,415 records uniquely to the located-event catalogue, using
identical coordinates after conversion from metres to millimetres and
chronological order. Unique locations anchor a mapping between the two
catalogue clocks, resolving repeated locations. The original event indices
retain the existing 23 hydraulic boundaries; the supplied ``Step'' labels
are not substituted for them. Of the matched events, 8,398 precede the final
boundary and 17 follow it. Amplitude coverage is 85\% overall and
68--100\% by stage. No amplitudes are assigned to missing records; weighted
and unweighted comparisons below use exactly the same available events.

Let $w_{ie}=\int_{V_i}G_e\,\mathrm{d}V$ be the integral of event $e$'s
Gaussian kernel over cell $i$, using the same 0.8~mm width and quadratic
interpolation as the existing density field. For events $E_j$ between
measurements $j-1$ and $j$, define
\begin{equation}
\Delta\delta^{*}_{i,j}=\frac{1}{V_i}\sum_{e\in E_j}w_{ie},\qquad
\Delta a_{i,j}=\frac{1}{V_i}\sum_{e\in E_j}w_{ie}A_e,\qquad
\overline{M}_{i,j}=\frac{\sum_{e\in E_j}w_{ie}M_{\mathrm{AE},e}}
{\sum_{e\in E_j}w_{ie}}.
\end{equation}
Here $E_j$ contains only events with amplitudes. The first two metrics measure
recent unweighted and amplitude-weighted activity. The third measures mean
relative event size, separating it from the amount of activity; we report
it only where $\sum_{e\in E_j}w_{ie}\geq1$, avoiding averages over nearly
empty kernel tails. Changing the common amplitude unit multiplies
$\Delta a$ by a constant and shifts $\overline{M}$ by a constant, leaving
the reported rank correlations unchanged.

For the cumulative comparison in main-text Figure~9, we use
$\delta^{*}_{i,j}=\sum_{r=1}^{j}\Delta\delta^{*}_{i,r}$ and
$a_{i,j}=\sum_{r=1}^{j}\Delta a_{i,r}$, so that corrected linear
amplitudes $A_e=10^{M_{\mathrm{AE},e}}$ are accumulated.
The resulting amplitude densities are expressed in V~mm$^{-3}$.
The first stage includes all preceding catalogue events. Both columns use
the same events with amplitudes and the same spatial weights. At sample
scale, cumulative catalogue counts and amplitude sums are divided by the
sample volume. The transmissivities are unchanged from the hydraulic
inversion; only the AE coordinates differ between columns.

Table~\ref{tab:si_ae_amplitude_metrics} compares the interval metrics with
$D_{i,j}=\log_{10}(T_{i,j}/T_{i,j-1})$. Amplitude weighting retains the
cell-dependent associations seen with counts. Excluding the three
intervals spanning overnight holds gives correlations of $-0.24$ to 0.54.
Repeating the amplitude-weighted comparison with the alternative
initialisation and off-fault permeability gives ranges of $-0.20$ to 0.46
and $-0.25$ to 0.37, respectively.
Sample-wide, the amplitude sum and mean $M_{\mathrm{AE}}$ over each interval
have correlations of $-0.01$ and $-0.04$ with the logarithmic change in mean
transmissivity ($n=22$). Cumulative amplitude and cumulative count are both
strictly increasing and therefore give the same rank correlation with mean
transmissivity (0.81); these are not independent evidence for an event-size
relation. The comparisons describe the available amplitude subset, with the
same within-experiment scope as Text S4.

\clearpage
\begin{table}[!ht]
\caption{Spearman correlations with local logarithmic transmissivity change
$D_{i,j}$. Count and amplitude increments use the same events and all 22
intervals; ``Amp., exc.'' excludes the three intervals spanning overnight holds
($n=19$). Mean relative magnitude $\overline{M}$ uses $n_M$ intervals with
at least one kernel-equivalent event in the cell. These coefficients describe
local associations, not independent predictive validation.}
\label{tab:si_ae_amplitude_metrics}
\centering
\small
\begin{tabular}{lrrrrr}
\toprule
Cell & Count & Amplitude & Amp., exc. & Mean $M_{\mathrm{AE}}$ & $n_M$ \\
\midrule
$C_{1}$ & 0.42 & 0.38 & 0.39 & 0.11 & 16 \\
$C_{2}$ & -0.13 & -0.21 & -0.21 & -0.37 & 21 \\
$C_{3}$ & 0.00 & -0.25 & -0.24 & -0.63 & 20 \\
$C_{4}$ & -0.34 & -0.12 & 0.15 & 0.40 & 21 \\
$C_{5}$ & 0.00 & 0.09 & 0.34 & -0.31 & 13 \\
$C_{6}$ & -0.22 & -0.14 & -0.10 & -0.19 & 13 \\
$C_{7}$ & 0.20 & 0.17 & 0.16 & -0.17 & 20 \\
$C_{8}$ & -0.01 & -0.13 & 0.00 & -0.17 & 20 \\
$C_{9}$ & 0.24 & 0.19 & 0.54 & 0.00 & 21 \\
$C_{10}$ & 0.04 & -0.12 & 0.09 & -0.45 & 18 \\%
\bottomrule
\end{tabular}
\end{table}